\documentclass[a4paper,fleqn,usenatbib]{mnras}

\usepackage{newtxtext,newtxmath}
\usepackage{float}
\usepackage{subcaption}

\usepackage[T1]{fontenc}
\usepackage{ae,aecompl}

\usepackage{graphicx}	% Including figure files
\usepackage{amsmath}	% Advanced maths commands
\usepackage{amssymb}	% Extra maths symbols
\usepackage{multicol}

\title[Interpreting ML methods for source classification]{Learning from the machine: interpreting machine learning algorithms for point- and extended- source classification}

\author[X. Morice-Atkinson et al.]{
Xan Morice-Atkinson$^{1}$\thanks{E-mail: \href{mailto:xan.morice-atkinson@port.ac.uk}{xan.morice-atkinson@port.ac.uk}},
Ben Hoyle$^{2,3}$,
David Bacon$^{1}$
\\
$^{1}$Institute of Cosmology and Gravitation, University of Portsmouth, Burnaby Rd, Portsmouth, PO1 3FX, UK\\
$^{2}$Ludwig-Maximilians-Universit\"at M\"unchen, Universit\"ats-Sternwarte, Scheinerstr. 1, D-81679 Munich, Germany\\
$^{3}$Max Planck Institute for Extraterrestrial Physics, Giessenbachstrasse, 85748 Garching, Germany
}

\date{Accepted XXX. Received YYY; in original form ZZZ}

\pubyear{2017}

\begin{document}
\label{firstpage}
\pagerange{\pageref{firstpage}--\pageref{lastpage}}

\maketitle

% Abstract of the paper
\begin{abstract}
We investigate star-galaxy classification for astronomical surveys in the context of four methods enabling the interpretation of black-box machine learning systems. 
The first is outputting and exploring the decision boundaries as given by decision tree based methods, which enables the visualization of the classification categories. 
Secondly, we investigate how the Mutual Information based Transductive
Feature Selection (MINT) algorithm can be used to perform feature pre-selection. If one would like to provide only a small number of input features to a machine learning classification algorithm, feature pre-selection provides a method to determine which of the many possible input properties should be selected. Third is the use of the \textit{tree-interpreter} package to enable popular decision tree based ensemble methods to be opened, visualized, and understood. This is done by additional analysis of the tree based model, determining not only which features are important to the model, but how important a feature is for a particular classification given its value. 
Lastly, we use decision boundaries from the model to revise an already existing method of classification, essentially asking the tree based method where decision boundaries are best placed and defining a new classification method. 

We showcase these techniques by applying them to the problem of star-galaxy separation using data from the Sloan Digital Sky Survey (hereafter SDSS). 
We use the output of MINT and the ensemble methods to demonstrate how more complex decision boundaries improve star-galaxy classification accuracy over the standard SDSS \texttt{frames} approach (reducing misclassifications by up to $\approx33\%$). We then show how \textit{tree-interpreter} can be used to explore how relevant each photometric feature is when making a classification on an object by object basis. 

\end{abstract}


%%%%%%%%%%%%%%%%%%%%%%%%%%%%%%%%%%%%%%%%%%%%%%%%%%

%%%%%%%%%%%%%%%%% BODY OF PAPER %%%%%%%%%%%%%%%%%%

\section{Introduction}

An important and long-standing problem in astronomy is that of object classification; for example, whether an object in a photographic plate is a nearby star or a distant galaxy. Independent of the data-sample under investigation, the process of building a source catalog will require object classification. There are multiple ways of determining the classification of astronomical objects, each with their own advantages and disadvantages. For example, template fitting methods applied to photometric or spectroscopic data can be accurate but are dependent on the choice of templates; whereas classifying objects by radial profile can be quick, but of limited accuracy due to the small amount of information used for each object. For instance, radial profile data alone cannot be used to distinguish between point sources, such as stars and QSOs. 

There are more complex methods of object classification that exist to identify astronomical objects, such as Artificial Neural Networks that use photometry to isolate high redshift QSOs \citep{2010Yeche}, or objects at fainter magnitudes \citep{2013Soumagnac}. There are also successful complex non-machine learning methods in use, such as likelihood functions \citep{2011Kirkpatrick} for point source separation, where an object is classified as a QSO based on the summed gaussian distance to every object in a set of known QSOs and stars in colour space.

This paper aims to introduce a new combination of machine learning data analysis methods to astronomy\footnote{Our code is hosted at https://github.com/xangma/ML\_RF}, specifically with the use case of object classification, although we note that these methods can readily applied to other problems. The goal is to use machine learning to improve the precision/purity of object classification from photometric data, while simultaneously analyzing the generated machine learning models in an effort to understand the decision making processes involved.

We achieve this by selecting data properties relevant to the classification problem, then using those data with a range of machine learning algorithms to classify astronomical objects. During the classification process, decision-making information that is usually internal to the machine learning algorithm will be gathered, output, and visualized to achieve a deeper understanding of how the machine learning algorithm succeeds in classifying individual objects.

The paper is laid out as follows. Section \ref{sec:data} describes the SDSS data and standard \texttt{frames} algorithm, Section \ref{sec:methods} describes the new methods employed in this work, including feature selection, a comparison of algorithm performance, and methods to interpret the decision making processes in one of the tree-based algorithms. Section \ref{sec:results} details the results obtained from these methods in terms of purity and completeness. In Section \ref{sec:discussion} we discuss the results and conclude.

\section{Data}
\label{sec:data}    
In this Section we introduce the observational data used in this paper, which is drawn from the Sloan Digital Sky Survey \citep[hereafter SDSS][]{1998Gunn}. We briefly review the standard photometric star/galaxy classification criterion given by the {\tt frames} method which is obtained through the query of the flag {\tt objc\_type} \citep{2002SDSS} in the CasJobs SkyServer \citep{2002Szalay}.

\subsection{Observational data}
The data in this work is drawn from SDSS Data Release 12 \citep[DR12,][]{2015DR12}. The SDSS  uses a 2.5 meter telescope at Apache Point Observatory in New Mexico and has CCD wide field photometry in 5 bands \citep[$u,g,r,i,z$][]{2010Doi,2006Gunn,2002Smith,2000York}, as well as an expansive spectroscopic follow-up program \citep[][]{2013Dawson,2011Eisenstein,2013Smee} covering $\pi$ steradians of the northern and equatorial sky. The SDSS collaboration has obtained more than 3 million spectra of astronomical objects using dual fiber-fed spectrographs. An automated photometric pipeline performs object classification to an $r$ band magnitude of $r\approx22$ and measures photometric properties of more than 100 million galaxies. The complete data sample, and many derived catalogs including galaxy photometric properties, are publicly available through the {\tt CasJobs} server \citep[][]{2008Li}\footnote{skyserver.sdss3.org/CasJobs}.

As we will draw large random samples from the SDSS DR12 data, we first obtain the full relevant dataset.
We obtain object IDs, magnitudes and errors as measured in different apertures in each band, radial profiles, both photometric and spectroscopic type classifications, and photometry quality flags using the query submitted to CasJobs shown in Appendix \ref{app:sql}. The resulting catalog is similar to that of the feature importance work of \citet{2015Hoyle}, but we omit redshift information. We generate a range of standard colors (e.g.,  \texttt{PSFMAG\_U-PSFMAG\_G}) and non-standard colors (e.g. \texttt{PSFMAG\_U-CMODELMAG\_G}) for each object. The final catalog contains 215 input quantities, or `features'. The magnitudes used in this work are corrected for galactic xtion where appropriate. We further select objects to have a clean spectroscopic classification by setting the \texttt{Zwarning} flag to be equal to 0. This selection removes $\approx11\%$ of the sample.  

The final catalog contains 3,751,496 objects. We note that approximately 66\% of these objects are spectroscopically classified as galaxies with the remaining objects classified as point sources. 
We select two random samples from the final catalogue: the first is a training sample of 10,000 objects and the second is a test sample comprised of 1.5 million objects. The small training sample allows a large exploration of model space to be completed in a tractable time scale.

\subsection{Existing SDSS Classification Schemes: Spectral Fitting and Photometric Selection} \label{sec:data_fr}
The SDSS provides both a spectroscopic and a photometric classification for each object which both attempt to infer if the object is a galaxy or a point source, including both stars and QSOs. We briefly review both techniques below.

The spectroscopic classification method is called \texttt{CLASS}, and compares spectral templates and the observed spectra using a $\chi^2$ cost function \citep{2012Bolton}. During this process galaxy templates are restricted between redshifts, $0<z<2$ and QSO templates are restricted to $z<7$. We note that the observed spectra are masked outside the wavelength range of 3600\AA to 10400\AA. This paper assumes that this analysis produces the true object classification, and we will use it to compare different photometric classification predictions. 

A second empirical method using photometric data is called \texttt{frames}, and uses the combination of following photometric magnitude measurements \texttt{PSFMAG-X - CMODELMAG-X}. The \texttt{PSFMAG} magnitude is calculated by fitting a point spread function model to the object which is then aperture corrected, as appropriate for isolated stars and point sources \citep[see][]{2002SDSS}. The \texttt{CMODELMAG} magnitude is a composite measurement generated by a linear combination of the best fit exponential and de Vaucouleurs light profile fits in each band. The resulting \texttt{CMODELMAG} magnitude has excellent agreement with Petrosian magnitudes for galaxies, and PSF magnitudes of stars \citep{2004SDSS}. Therefore the condition \texttt{PSFMAG-X - CMODELMAG-X} is a reasonable discriminator between galaxies and point sources.

In detail the composite feature \texttt{PSFMAG-X - CMODELMAG-X} is divided into two bins for each of the X=5 SDSS bands, and the separating condition used to determine the object class is the same for each band and given by
\begin{equation}
\texttt{PSFMAG} - \texttt{CMODELMAG} > 0.145 $\,$.
\label{eq:frames}
\end{equation}

The SDSS pipeline provides the \texttt{frames} classification for each object in each photometric band, as well as an overall classification calculated by summing the fluxes in all bands and applying the same criterion as in Equation \ref{eq:frames}. We use this latter summation as the base line SDSS photometric classification scheme in this work.

\begin{figure}
\includegraphics[width=\columnwidth]{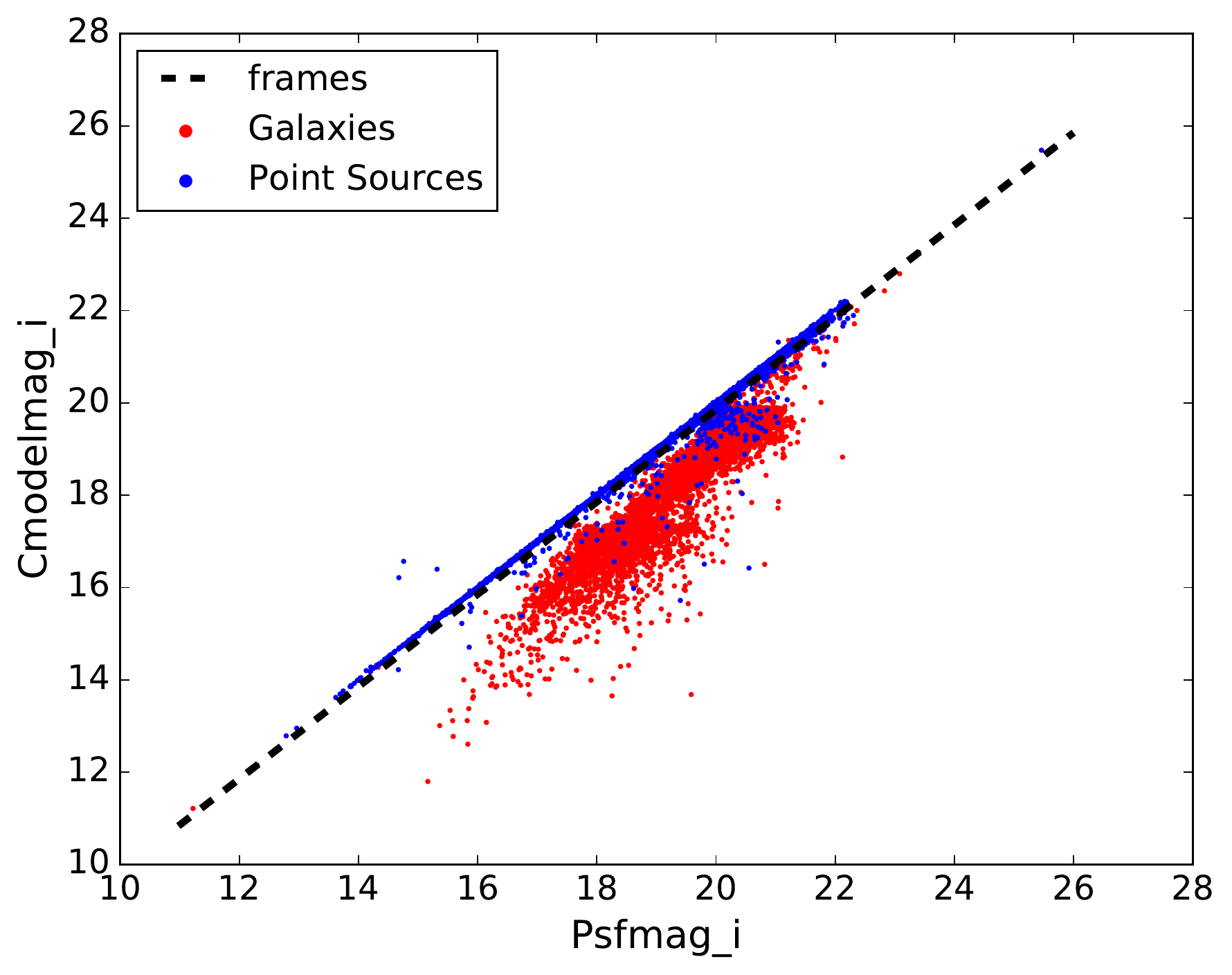}
\caption{Object classification using the \texttt{frames} method. Here we show the relevant difference between two magnitude estimates in the I band, with the discriminating dashed black line drawn according to Equation \ref{eq:frames}.} 
\label{fig:frames_cut}
\end{figure}

We show the distribution in \texttt{PSFMAG} and \texttt{PSFMAG} - \texttt{CMODELMAG} for the training sample in Figure  \ref{fig:frames_cut}, and show the condition given in Equation \ref{eq:frames} by the solid black line.

In this paper we investigate if a new photometric classification can improve the accuracy of the \texttt{frames} methods, and if by understanding how some machine learning systems work, we can motivate changes to these base-line photometric classification schemes. The authors of the \texttt{frames} method state that it works at the 95\% confidence level to $r=$21, and that the method becomes unreliable at fainter magnitudes \citep{2002SDSS}.

\subsection{Data Preparation} \label{sec:datprep}
For the main body of this work, we only select data with good photometry and spectra. In particular we select \texttt{clean} flag $=1$. This removes objects which are duplicates, or with deblending issues, interpolation issues, or have suspicious detections, or are stars close to the edge of the survey\footnote{see http://www.sdss.org/dr12/algorithms/photo\_flags\_recommend/}.

We explore how this may bias our results, and perform a standalone test in Section \ref{sec:results_clean} with and without the \texttt{clean} flag selection to determine what effect this has on our accuracy.

\section{Methods}
\label{sec:methods}
This section introduces the machine learning algorithms used in this work, including the methodologies behind the MINT feature selection algorithm \citep{2013He} and a method to simplify ensemble methods based on decision trees called \textit{treeinterpreter} \citep{2015andosa}. We also describe how machine learning algorithms can be used to motivate improvements to the base-line SDSS \texttt{frames} classification.

\subsection{Object Classification Using Machine Learning Methods} \label{sec:methods_ml}
Four tree based machine learning methods are used in this work; Random Forest \citep[RF,][]{2001Breiman}, Adaboost \citep[ADA,][]{1997Freund, 2009Zou}, Extra Randomised Trees \citep[EXT,][]{2006Geurts} and Gradient Boosted Trees \citep[GBT,][]{1999Friedman, 2001friedman, 2009Hastie}. We use the implementations of these algorithms from within the \texttt{scikit-learn} package. All of these methods are able to draw a decision boundary in multidimensional parameter spaces which distinguishes classification classes. We describe these algorithms briefly below.

A decision tree is a flowchart-like model that makes ever finer partitions of the input features (here photometric properties) of the training data. Each partition is represented by a branch of the tree. The input feature and feature value used to generate the partitions are chosen to maximise the success rate of the target values (here point source or galaxy classifications) which reside on each branch. 
This process ends at leaf nodes, upon which one or more of the data sit. A new object is queried down the tree and lands on a final leaf node. It is assigned a predicted target value from the true target values of the training data on the leaf node. A single decision tree is very prone to over fitting training data.

Random Forests train by generating a large number of decision trees, with each tree using a bootstrap re-sample of the training data and a random sample of the input features. During classification of new data the majority vote across all trees is taken. By building a model that takes a vote from many decision trees, the problem of over fitting the training set is overcome, allowing better generalization to unseen data. 

Extra Randomized Trees is a similar algorithm to Random Forests, but splits in the generated decision trees are decided at random instead of calculating a metric. This makes model training faster and can further improve generalization. 

Adaboost and Gradient Boosted Trees are both examples of boosted algorithms, which convert so-called decision stumps into strong learners. Decision stumps are shallow decision trees that result in predictions close to a random guess. The data is processed through these trees multiple times with the algorithm weighting the model based on performance. Adaboost changes the model between iterations by re-weighting the data of objects that were misclassified at a rate governed by the \textbf{learning\_rate} parameter. This minimises model error by focusing the subsequent tree on those misclassified objects.
Gradient Boosted Trees changes the model by iteratively adding decision stumps according to the minimisation of a differentiable loss function (which tracks misclassification) using gradient descent.
The model will start with an ensemble of decision stumps and the loss will be assessed. Between each iteration the algorithm adds decision stumps that reduce the loss of the model, stopping when loss can no longer be reduced (when the gradient of reducing loss flattens).

In this paper, we will perform object classification using each of these four algorithms, for each of the following three subsets of photometric features: 
\begin{itemize}
  \item the five features that the SDSS pipeline uses in the \texttt{frames} method (i.e., \texttt{PSFMAG} - \texttt{CMODELMAG} for each filter);
  \item five features selected using a feature selection method, MINT as discussed in Section \ref{subs:MINT};
  \item all 215 features available in the sample.
\end{itemize}
Each test is performed with 10000 objects in the training sample, predicting on a test sample of 1.5 million objects. We will show the results for accuracy of classification in Section \ref{sec:results_ml} for each algorithm (Random Forests, Extra Randomised Trees, Gradient Boosted Trees, and Adaboost), operating on the different subsets of photometric features.

\begin{figure}
\includegraphics[width=\columnwidth]{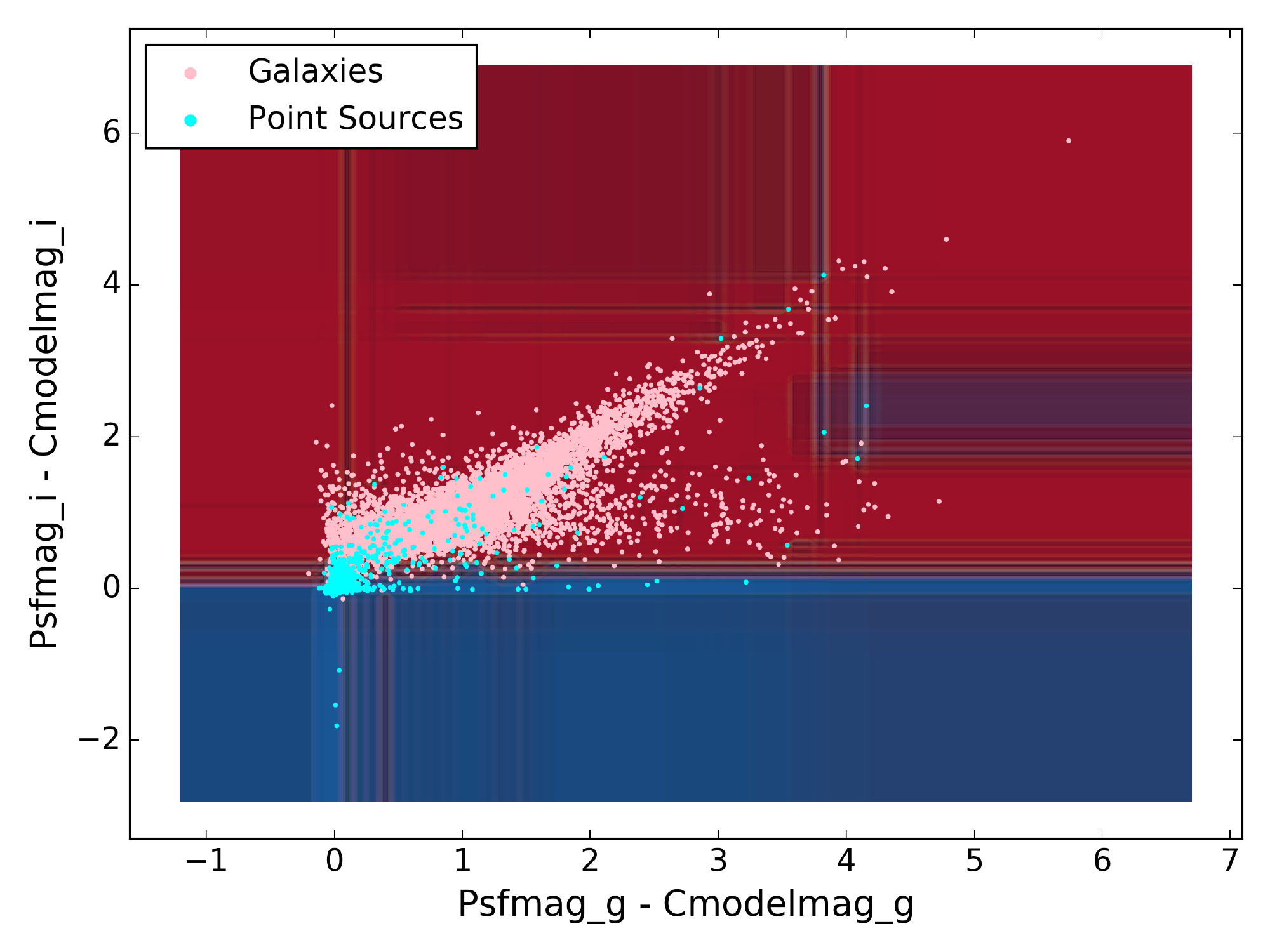}
\caption{Training data (pink and cyan points for galaxies and point sources) plotted over the decision boundaries (red and blue background for galaxies and point sources), generated by an example Random Forest run using \texttt{frames} features in g and i band. The colour of the training data denotes spectroscopic classification.}
\label{fig:db_cut}
\end{figure}

Figure \ref{fig:db_cut} shows an example of the decision boundaries created from a Random Forest run using only two features, a simplified version of the first test in the list above. The area where the algorithm classifies objects as galaxies is shown in red, with classifications of stars shown in blue. The areas where classifications are more distinct have bolder colours, with the area around the horizontal boundary showing more uncertainty in object classification. The plotted points show all 10000 objects of the training sample, colour-coded by their spectroscopic type. It should be noted that the Random Forest draws boundaries very similar to the ones in the SDSS pipeline paper, though not as linear. However, it can be seen that some objects are misclassified using both the \texttt{frames} method and this particular Random Forest run. Using more than two features, such as in the tests listed above, allows the machine learning methods to utilize more dimensions in parameter space and consequently achieve a higher accuracy of classification. \\

\subsection{Feature selection using MINT}
\label{subs:MINT}
The SDSS pipeline measures and calculates a rich abundance of features from the photometric images. Rather than just focusing on those features employed in the \texttt{frames} algorithm, one may also choose other available features to pass to the machine learning algorithms. To aid in the interpretation of the results it would be advantageous to select only a small number of features, but chosen wisely such that they are minimally correlated with each other, and have strong predictive power.

A suitable method of feature selection is `Maximum Relevance and Minimum Redundancy' (mRMR) which can help to find a small number of relevant input features without relinquishing predictive power. This has been proven to work in multiple datasets involving e.g. handwritten digits, arrhythmia, NCI cancer cell lines, and lymphoma tissues \citep{2005Peng, 2005Peng2}. 

mRMR first calculates the maximum relevance, a feature selection method based on the measurement of mutual dependence (correlation) between the variables. In the case of this work, the variables are the features for each object (e.g. \texttt{CMODELMAG\_G}), and the class is galaxy or point source. Maximum relevance measures the mean of all of the mutual information values (a measure of correlation) between unique pairs of individual features $x_{i}$, and classes $c$, with the aim of finding a set of features most correlated with a specific classification. The maximum relevance calculation is

\begin{equation}
max\, D(S,c)\\
D=\frac{1}{\left | S \right |} \sum_{x_{i}\in S}^{ } I(x_{i};c),
\label{eq:maxR}
\end{equation}

where $D$ is maximised for the selected features $S$ and class $c$, with $I$ being the mutual information. 
Selecting features that are maximally relevant to the classification causes the set of returned features to be highly correlated with one another. To compensate for this, features that are highly correlated with other features are removed using minimum redundancy which is calculated as

\begin{equation}
min\, R(S)\\
R=\frac{1}{\left | S \right |^{2}} \sum_{x_{i},x_{j}\in S}^{ } I(x_{i},x_{j}),
\label{eq:minR}
\end{equation}
where $I$($x_i$,$x_j$) represents the mutual information between features $x_i$ and $x_j$, and $R$ is minimised for the selected features $S$.

We would like to maximise $D$ while minimizing $R$. This can be simplified, completing the mRMR calculation by combining these requirements in one equation, and maximizing $\Phi$ where

\begin{equation}
max\, \Phi (D,R)\\
\Phi=D-R.
\label{eq:mRMR}
\end{equation}

This ensures the returned set of selected features is highly correlated with the classification, but are mutually exclusive from other features in the set. 

This work uses an extension of mRMR called Mutual Information based Transductive
Feature Selection (MINT) \citep{2013He}, a method designed to help with the `curse of dimensionality' in genome trait prediction. This arises due to the issue of having many more features than samples in the dataset. MINT assesses the mutual information between the training sample's features and classification and, setting it apart from mRMR, between individual features in both the training and test sample.

This means that MINT can effectively combine Equations \ref{eq:maxR} and \ref{eq:minR} into Equation \ref{eq:mRMR}, the same as mRMR, but is able to exploit a much larger amount of data due to the assessment of the correlation between features for the entire sample, not just the training sample. For this work, MINT is able to utilize photometric data from the 1.5 million objects in the test sample. This allows us to ensure, much more than we could using mRMR alone, that the selected features will be those which are correlated least with one another, thus giving us the best chance of accurate object classification. 

We will now consider an expanded version of Equation \ref{eq:mRMR} with the MINT modifications included. The incremental search for features using MINT works in the following way; We assume we have a set of $X$ total features, and $S_{m-1}$ as a subset of those features containing $m-1$ features. The m-th feature is selected from the remaining feature set, $X-S_{m-1}$, by maximising $\Phi$ in the same way as in Equation \ref{eq:mRMR}, as follows: 

\begin{equation}
\begin{split}
&max_{_{x_{j}}\in X - S_{m-1}}\\
&[I(x_{j}^{\texttt{Tr}};c^{\texttt{Tr}})-\frac{1}{m-1} \sum_{x_{i} \in S_{m-1}}^{ } I(x_{j}^{\texttt{Tr + Test}};x_{i}^{\texttt{Tr + Test}})].
\end{split}
\label{eq:MINT}
\end{equation}

The modifications are made clear by the indication of which sample set is being used in the mutual information calculations, either only the training (\texttt{Tr}), or both training and test (\texttt{Test}) samples. 

We follow \cite{2013He} and explore the high dimensional feature space using the greedy algorithm \citep{2002Vince}.

In the case of MINT, greedy means that parts of the calculation are performed dynamically - utilizing previously calculated values in the MINT algorithm for future MINT calculations - making the feature selection process vastly quicker.

A user defined number of features is selected using the MINT algorithm, thus reducing the amount of input data (by reducing the number of features) required to make a robust prediction for the test sample. In this work, we reduce the number of features from 215 to 5 using MINT. This is to mirror the number of features the \texttt{frames} method uses and to test whether we can make accurate predictions with severely reduced data per object.

\begin{table}
\begin{center}
Number of selected MINT features (using 10000 training objects and 1.5 million test objects)
\begin{tabular}{ |c|c|c|c|c|c|c|c| }
 \hline
 5 &  10 \\ 
 \hline
 \texttt{PSFMAG\_G - CMODELMAG\_R} & 
 \texttt{DERED\_Z - FIBERMAG\_R} \\
 \texttt{PSFMAG\_I - FIBERMAG\_I} & 
 \texttt{PSFMAG\_I - CMODELMAG\_I} \\
 \texttt{DERED\_G - FIBERMAG\_G} & 
 \texttt{PSFMAG\_I - FIBERMAG\_I} \\
 \texttt{PSFMAG\_I - CMODELMAG\_I} & 
 \texttt{DERED\_G - FIBERMAG\_G} \\
 \texttt{PSFMAG\_R - FIBERMAG\_Z} & 
 \texttt{PSFMAG\_G - CMODELMAG\_R} \\
 & \texttt{PSFMAG\_Z - FIBERMAG\_Z} \\
 & \texttt{PSFMAG\_G - CMODELMAG\_G} \\ 
 & \texttt{PSFMAG\_R - FIBERMAG\_Z} \\
 & \texttt{DERED\_R - PSFMAG\_R} \\
 & \texttt{PSFMAG\_R - FIBERMAG\_R} \\
\end{tabular}
\caption{The features selected by MINT when setting the total number of features to five, or ten.}
\label{tab:nMINT}
\end{center}
\end{table}

Table \ref{tab:nMINT} shows the results of the MINT feature selection method for 5 or 10 total selected features. It can be seen that there are features in common between these two sets;  these have clearly been identified as robust and distinct features for classification.

\begin{figure}
\includegraphics[width=\columnwidth]{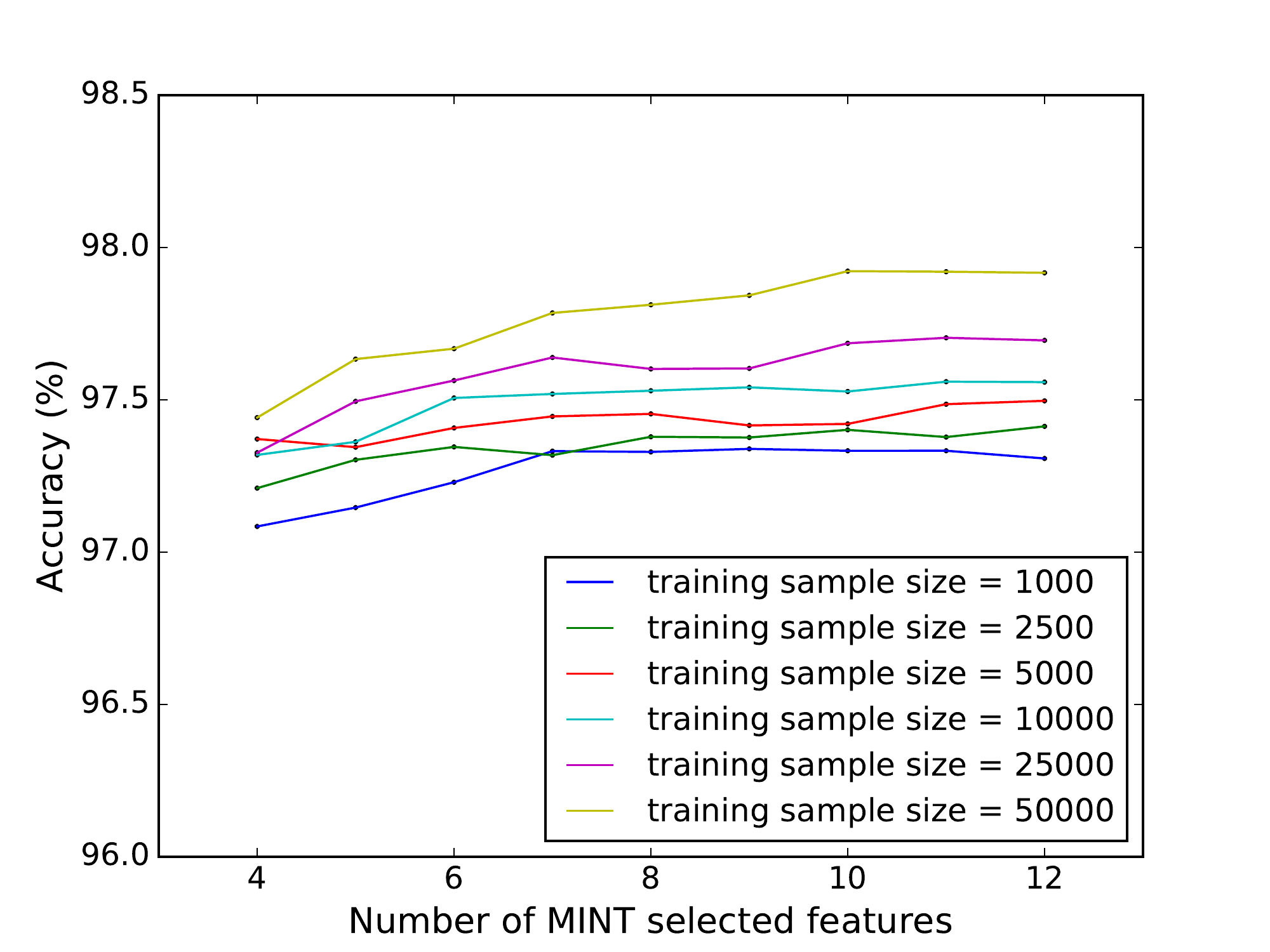}
\caption{Effect of number of MINT selected features on predictive accuracy. Coloured lines denote the number of objects used in the training sample.}
\label{fig:nMINT}
\end{figure}

We investigated the effect of changing the number of MINT selected features on the classification accuracy in a test Random Forest run (with 256 trees and no set maximum depth). This can be seen in Figure \ref{fig:nMINT}. The accuracy of the results only increases slightly ($\approx0.2$\%) as the number of MINT selected features increases. Also shown is the effect of changing the number of objects in the training sample. Again, the accuracy does not change significantly (<1\%).

\subsection{Interpreting Models of Tree Based Methods}
We use tree based machine learning methods due to their robustness, resistance to overfitting, and the ability to interpret them \citep{2009Hastie}. Understanding the workings of the model is most easily achieved by looking at the decision trees used by the algorithm. However, when the data are vast and complex and an ensemble of trees is used, the scope of the model deepens to such a degree that interpretation becomes near impossible. It is for this reason that new methods of model interpretation must be investigated.

\begin{figure*}
\includegraphics[width=\textwidth]{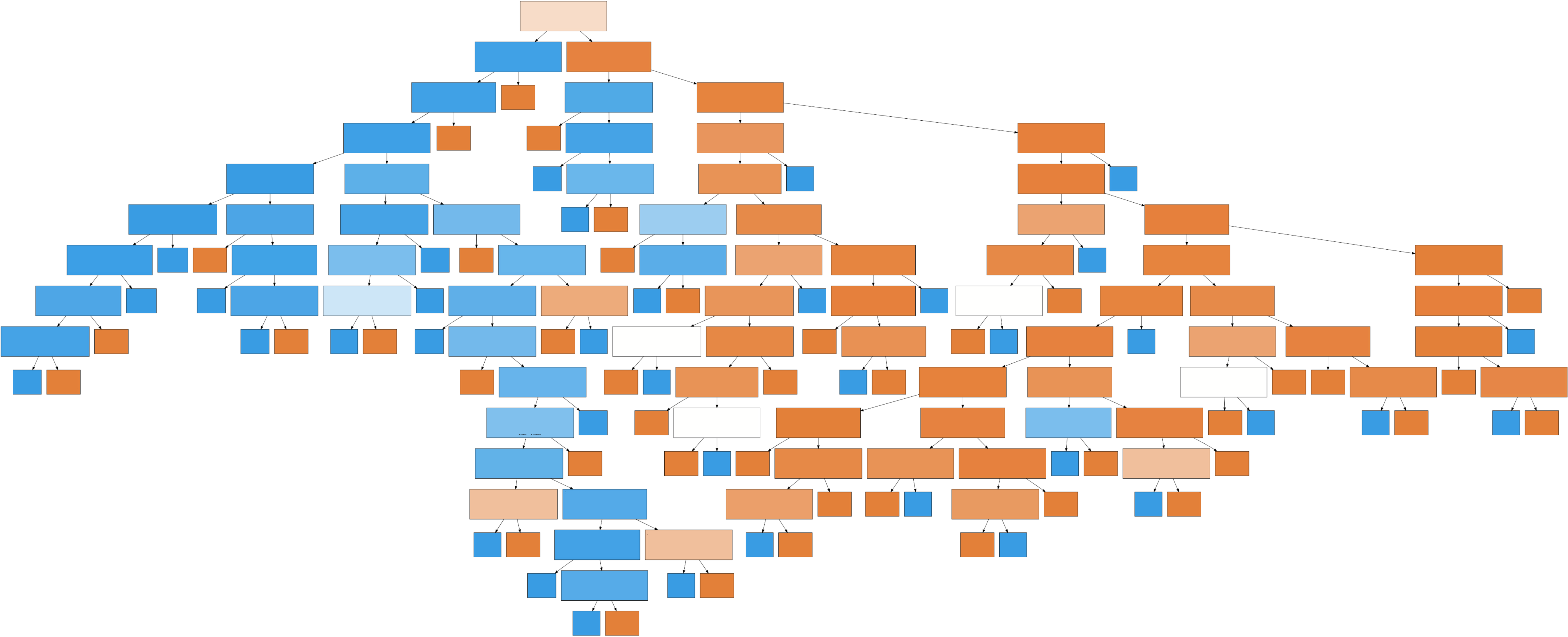}
\caption{Example of a single decision tree from a Random Forest comprised of 256 trees with unrestricted maximum depth. Blue colours indicate a point source classification, while orange colours indicate a galaxy classification. Opacity of colour represents probability of classification with more solid colours denoting higher probabilities.}
\label{fig:tree_ex}
\end{figure*}

An example decision tree taken from a Random Forest comprised of 256 trees and no limit on the hyperparameter ``maximum depth'', can be seen in Figure \ref{fig:tree_ex}. In this work, an example of how a node splits data (an example question in the tree) would be \texttt{PSFMAG\_R - CMODELMAG\_R} $\leq$ 0.25. Depending on the answer to this question, the object would advance through the tree in one direction or another towards the leaves (predicted class). It is clear from the complexity of the tree that it is unfeasible to easily gain information relating to the inner workings of the model by simply looking through the trees. This is especially the case since each tree will have drawn different decision boundaries relating to specific types of objects. For example, one tree may be very good at classifying red point sources, while another may excel at classifying blue galaxies.

There are methods for determining which features are important to the machine learning model, such as the  \texttt{feature\_importance} function provided in the scikit-learn package \citep{sklearn}. This is sometimes referred to as the ``mean decrease impurity", which is the total decrease in node impurity, an assessment of how well the model is splitting the data, averaged over all of the trees in the ensemble \citep{1984Breiman}. This is to say that the features in the model are assessed, and if they consistently contribute to making classifications, their importance increases.
This is useful, but somewhat ambiguous as it does not give much insight into the individual decisions the trees make, such as where it is most efficient to draw a boundary in parameter space. 

Instead, \textit{treeinterpreter}\footnote{\url{https://github.com/andosa/treeinterpreter}} \citep{2015andosa} can be used in an effort to decipher this information. For each object, \textit{treeinterpreter} follows the path through the tree, taking note of the value of the feature in question every time it contributes or detracts from an object being given a particular classification. This means that one can investigate how much the value of a particular feature contributes to the probability of a certain classification. 

To learn how \textit{treeinterpreter} works, we start with the mathematical description for a prediction given by a single tree.
The probability of a particular object being a member of class $c$ is given by the prediction function $f(x)$, where $x$ is the feature vector for the object in question. In the case where $f(x)$ is obtained from a single tree, we have

\begin{equation}
f(x)=c_{full} + \sum_{k=1}^{K}contrib(x,k),
\label{eq:TItree}
\end{equation}

where $c_{full}$ is the initial classification bias due to the class distribution in the sample for the class $c$, and $contrib(x,k)$ is the contribution from feature $k$ in the feature vector $x$ to the probability of being classified as class $c$. This means that the probability that the tested object is a galaxy from a single decision tree built using the whole training sample is a combination of two elements; the bias of galaxies (i.e. larger fraction) in the sample ($\approx66$\%) and the summation of the contribution to the probabilities the object was given due to the values of its features (the photometric quantities the object has) after the tree processed it. If there were no splits in the tree, the probability that any object in the test sample was a galaxy would remain at 66\%. 

Extending this to an ensemble of trees is fairly straightforward; the overall prediction function $F(x)$ from a Random Forest is the average of those of its trees $f_{j}(x)$,

\begin{equation}
F(x)=\frac{1}{J}\sum_{j=1}^{J}f_{j}(x),
\label{eq:TIRF1}
\end{equation}
where the number of trees is given as $J$. 

There is one last consideration to account for in the \textit{treeinterpreter} calculation; if each decision tree has been built using a bootstrap of the whole sample, the initial bias of the tree, $c_{full}$, will be different for each tree. It is for this reason that the bias terms of each tree are averaged and added to the average contribution of each feature. This makes the full equation in \texttt{treeinterpreter} for the prediction function 

\begin{equation}
F(x)=\frac{1}{J}\sum_{j=1}^{J}c_{j_{full}}+\sum_{k=1}^{K}(\frac{1}{J}\sum_{j=1}^{J}contrib_{j}(x,k)).
\label{eq:TIRF2}
\end{equation}

\begin{figure*}
\centering
\begin{subfigure}[t]{.45\linewidth}
\includegraphics[width=\columnwidth]{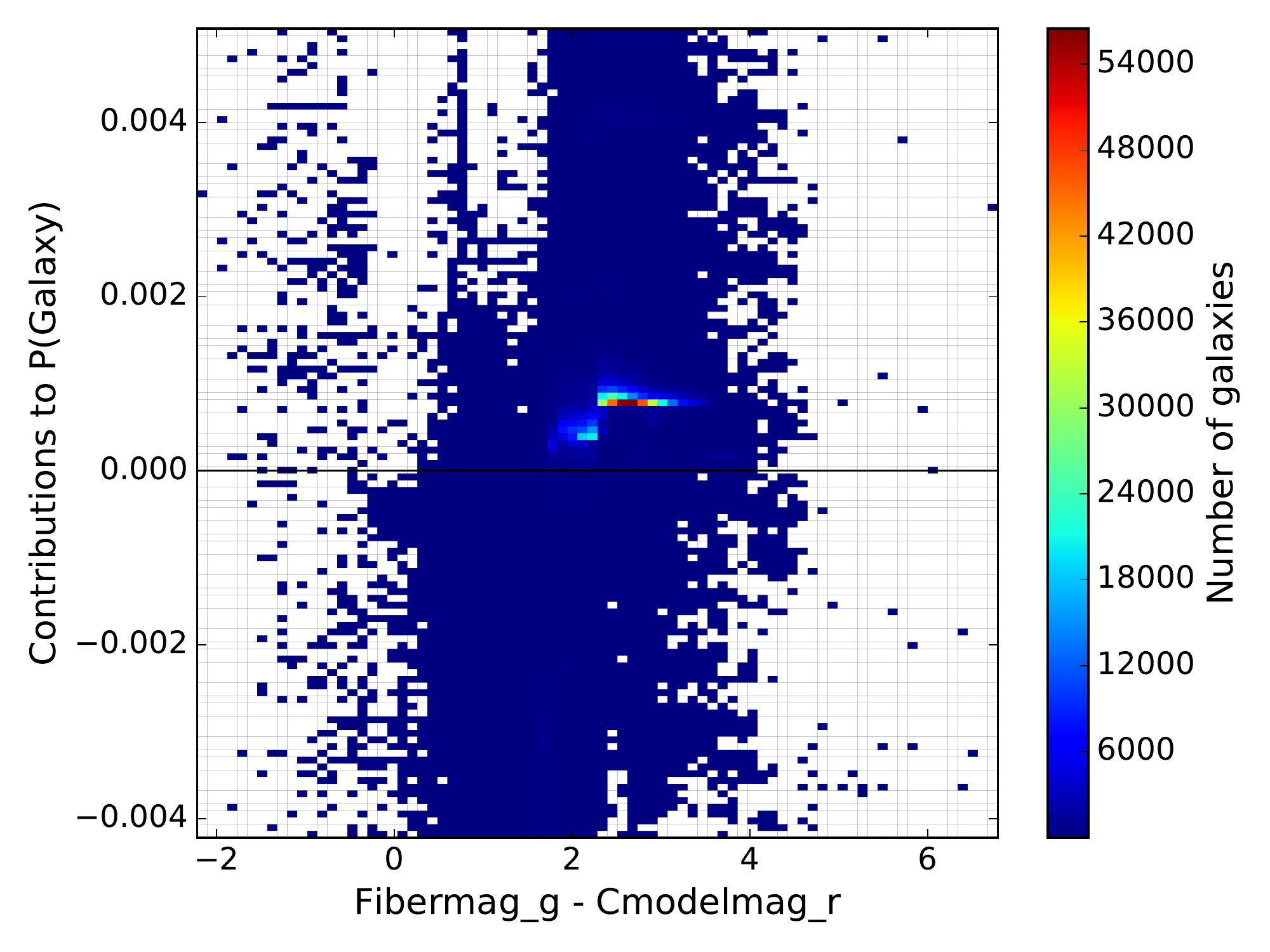}
\caption{The contribution to the probability of being predicted a galaxy by \texttt{FIBERMAG\_G - CMODELMAG\_R} of all spectroscopically confirmed galaxies in sample. Colour represents number of galaxies.}\label{fig:col_cont_a_}
\end{subfigure}\hfill
\begin{subfigure}[t]{.45\linewidth}
\includegraphics[width=\columnwidth]{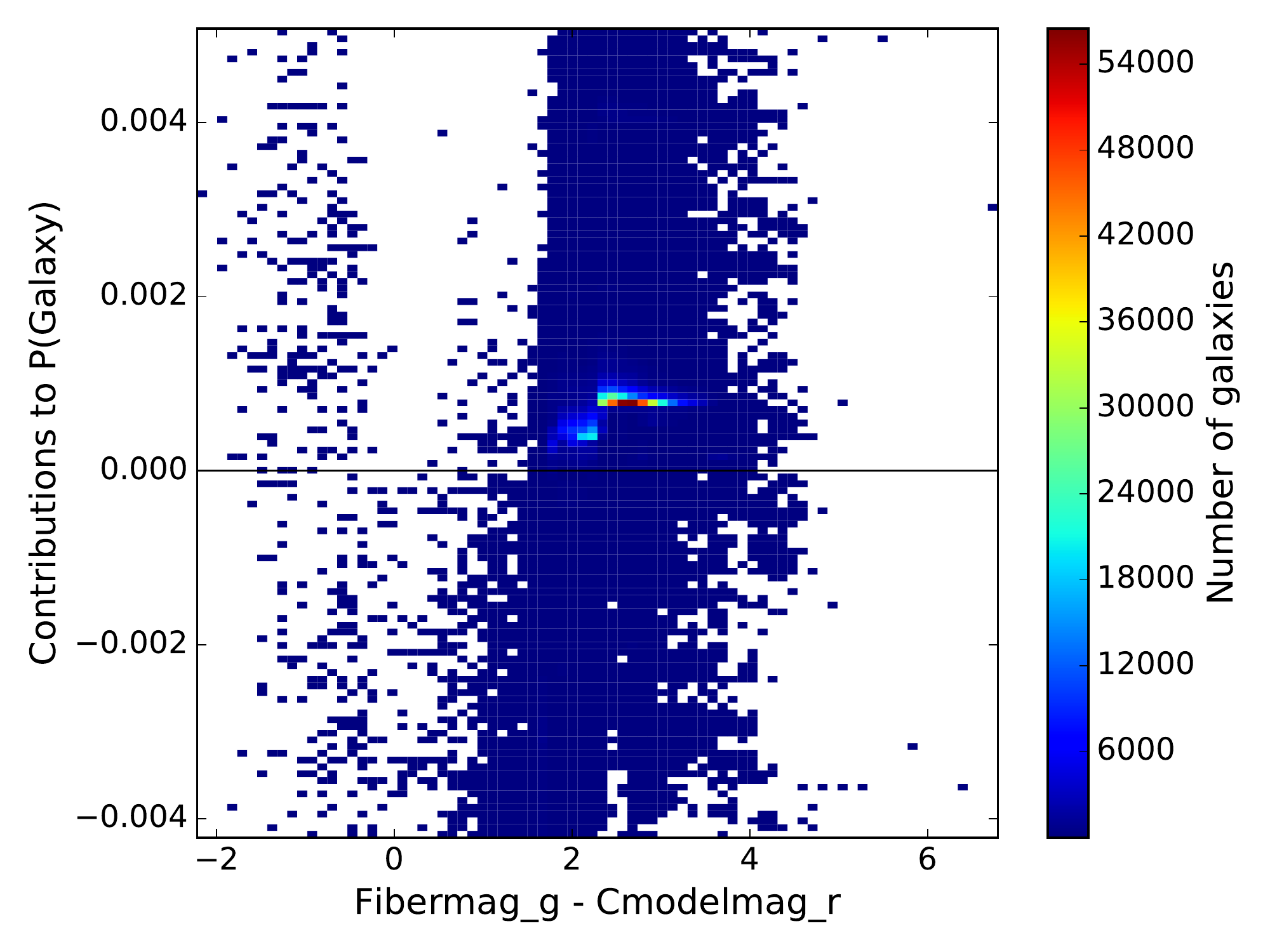}
\caption{The contribution to the probability of being predicted a galaxy by \texttt{FIBERMAG\_G - CMODELMAG\_R} for galaxies that have been correctly classified. Colour represents number of galaxies.}\label{fig:col_cont_b_}
\end{subfigure}\hfill
\begin{subfigure}[b]{.45\linewidth}
\includegraphics[width=\columnwidth]{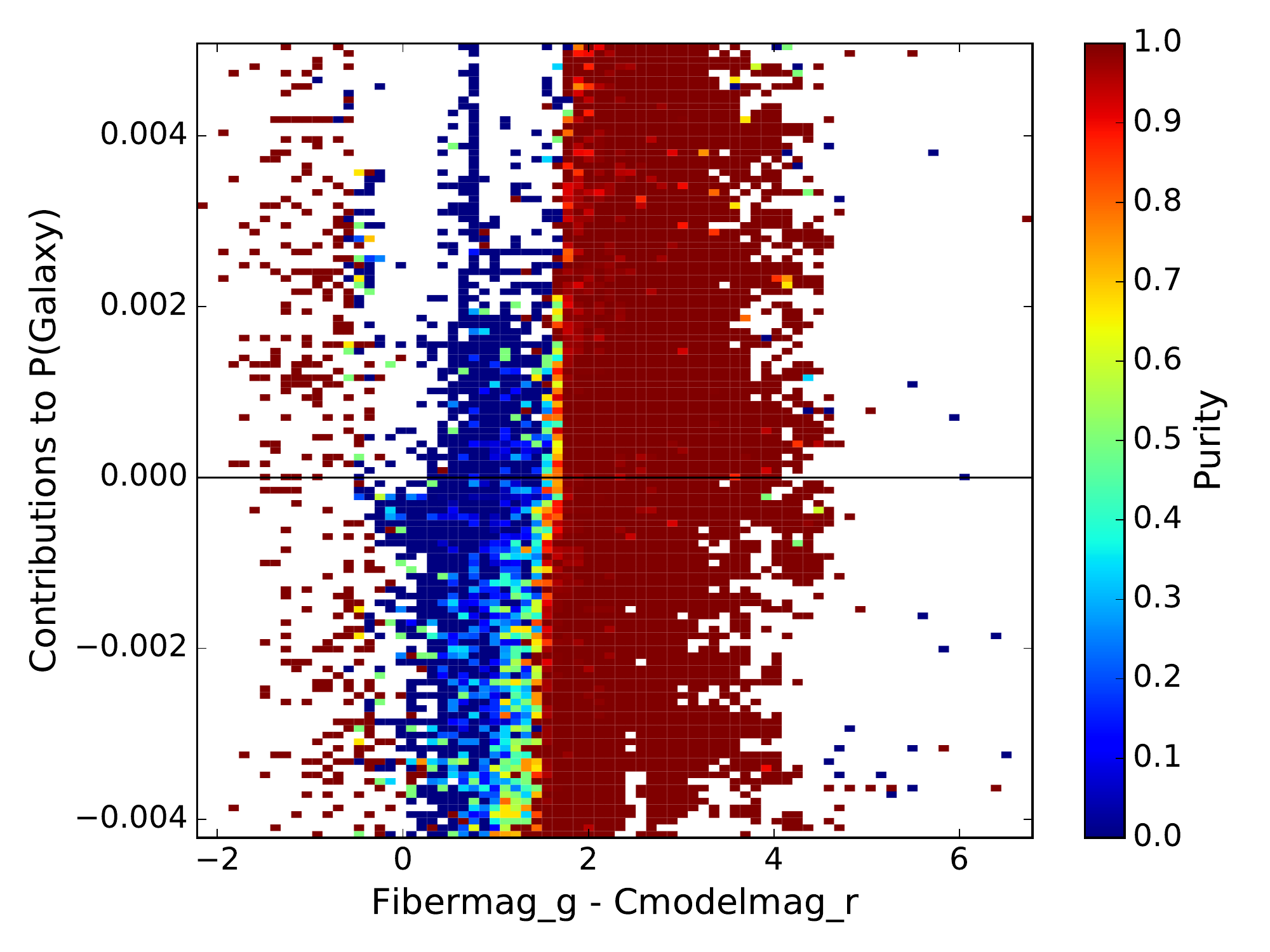}
\caption{The contribution to the probability of being predicted a galaxy by \texttt{FIBERMAG\_G - CMODELMAG\_R} for all objects classified as galaxies where the colour represents model purity.}\label{fig:col_cont_c_}
\end{subfigure}\hfill
\begin{subfigure}[b]{.45\linewidth}
\includegraphics[width=\columnwidth]{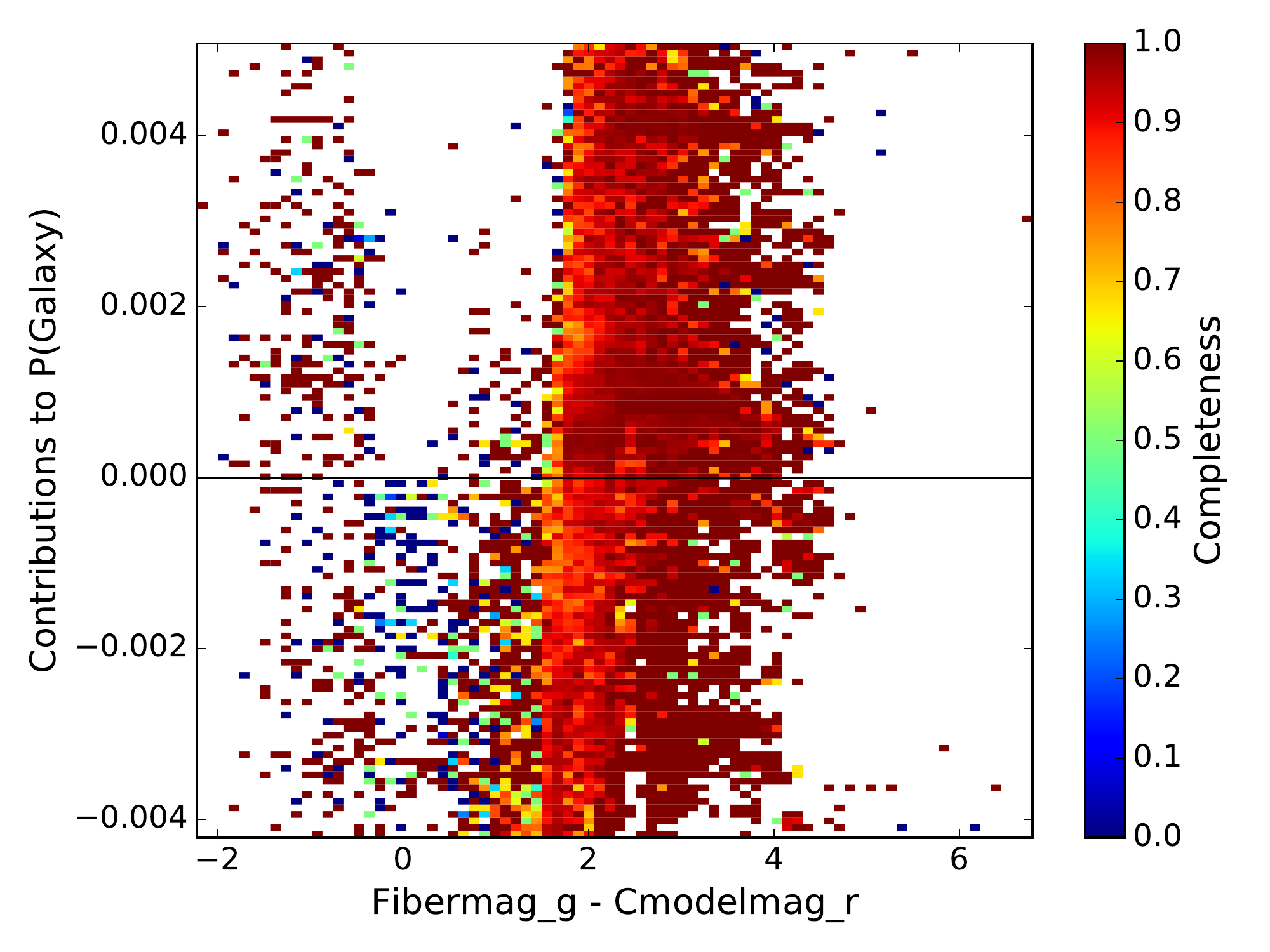}
\caption{The contribution to the probability of being predicted a galaxy by \texttt{FIBERMAG\_G - CMODELMAG\_R} for all spectroscopically confirmed galaxies where the colour represents model completeness.}\label{fig:col_cont_d_}
\end{subfigure}\hfill
\caption{Density plot of contributions to the probability of a galaxy classification by \texttt{PSFMAG\_G - CMODELMAG\_I} for spectroscopically confirmed galaxies. Purity refers to the fraction of retrieved instances that are relevant; completeness is the fraction of relevant instances that are retrieved. In relation to this work, purity would be a measure of how many galaxy classifications correctly identified galaxies, and completeness would be a measure of how many galaxies were correctly identified out of the total amount of galaxies. This example was created with a Random Forest comprising of 256 trees with no maximum depth, using all 215 available features.}
\label{fig:col_cont}
\end{figure*}

This not only presents which features are important to a particular classification in the model overall, but also which features were important for the individual classification of each object. As we know the value of the feature for each object, we can determine where in parameter space the model is succeeding or failing. This is visualized in Figure \ref{fig:col_cont}, where we present results for a particular example feature \texttt{FIBERMAG\_G - CMODELMAG\_R}; this feature's results exemplify several notable behaviours.

Figure \ref{fig:col_cont_a_} shows the contribution to the probability of galaxy classification from \texttt{FIBERMAG\_G - CMODELMAG\_R}, for all of the galaxies in the test sample, given a Random Forest model trained on 10000 objects (using 256 trees and all 215 features in our catalogue) . The colours show the number of objects with white showing the absence of data. Most of the galaxies fall into a small line of assigned probability of 0.002 at a \texttt{FIBERMAG\_G - CMODELMAG\_R} value of approximately 2.3, the mean of the sample, with the remaining galaxies scattered around the plot making up the blue colour. 
For this particular feature,\texttt{FIBERMAG\_G - CMODELMAG\_R}, some objects in the sample are given a reduced probability of being galaxies (i.e. they receive a negative contribution to probability); these are the data points below the black line. The model does not necessarily incorrectly classify these galaxies due to this one feature; there may be other features that are more important to the model than this one for classifying these particular galaxies.

Figure \ref{fig:col_cont_b_} shows the same as \ref{fig:col_cont_a_}, but for all the galaxies in the test sample that were correctly classified as galaxies. The colouring is the same as in Figure \ref{fig:col_cont_a_}. There are a number of galaxies with a \texttt{FIBERMAG\_G - CMODELMAG\_R} value of 0 to 2 that were incorrectly classified as point sources by the model, as they are missing when comparing to Figure \ref{fig:col_cont_a_}. 

\begin{figure}
\includegraphics[width=\columnwidth]{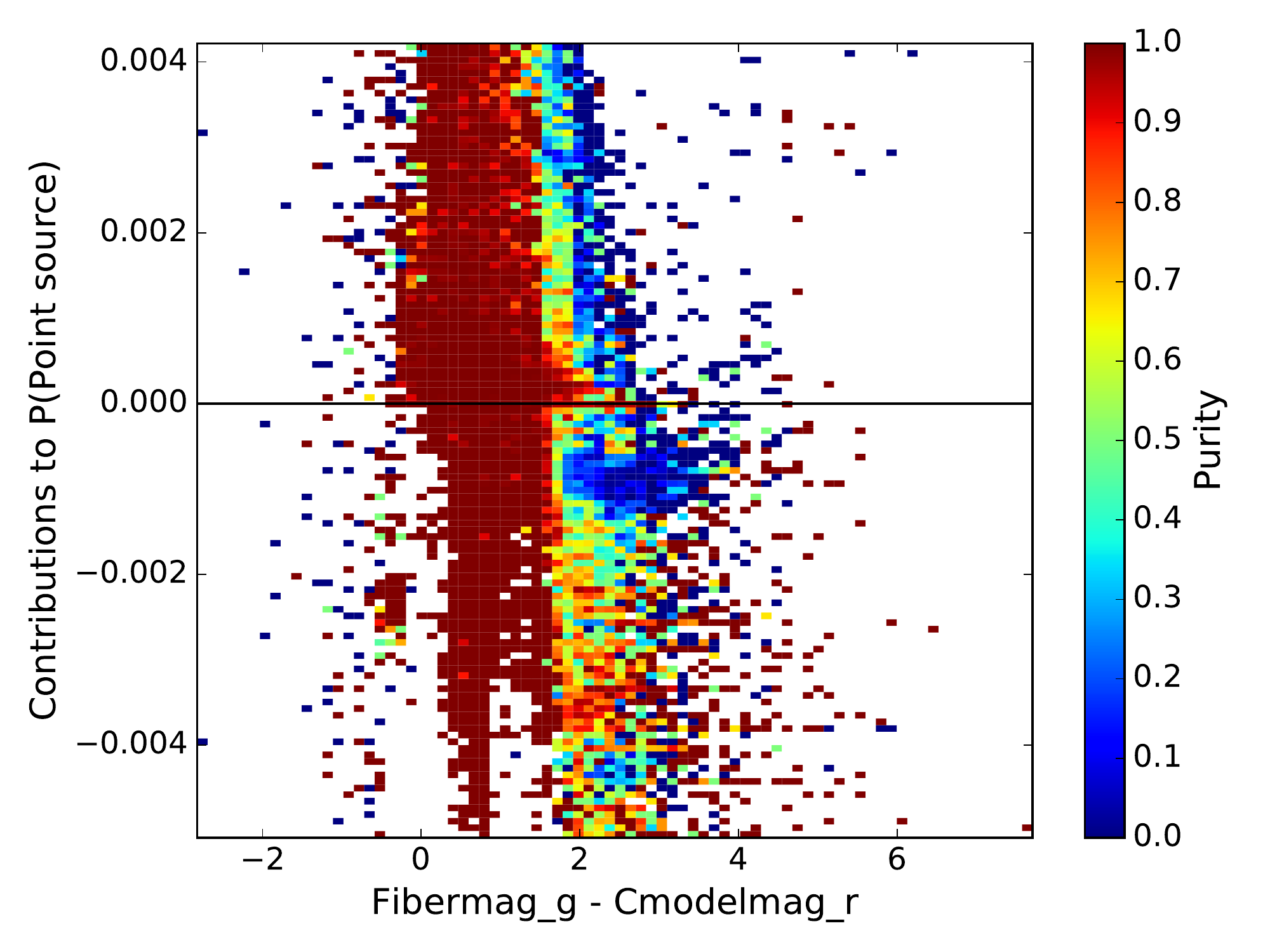}
\caption{The contribution to the probability of being classified as a point source by \texttt{FIBERMAG\_G - CMODELMAG\_R} where the colour represents purity. The correctly classified point sources here are occupying the parameter space of the incorrectly classified galaxies in Figure \ref{fig:col_cont_c_}.}\label{fig:col_cont_PS}
\end{figure}

The colour of Figure \ref{fig:col_cont_c_} shows the purity of the galaxy classification, the fraction of retrieved instances that are relevant. Here it can be seen that the model has failed to correctly classify bluer galaxies, where \texttt{FIBERMAG\_G - CMODELMAG\_R} is closer to 0. This is because that region of parameter space is being used to classify point sources, see Figure \ref{fig:col_cont_PS}.

The colour of Figure \ref{fig:col_cont_d_} shows the completeness of the galaxy classification; this can be interpreted as the probability that the object will be a galaxy given the model. Around values of \texttt{FIBERMAG\_G - CMODELMAG\_R} = 0, it can be seen that the model begins to fail at classifying galaxies correctly.

Visual analysis of this kind provides insight into how the model is drawing boundaries in parameter space, and information about where the classifications' limitations arise. 

\subsection{Performance of Algorithms}
Each machine learning method used in this work was tuned to optimise classification performance. This is achieved by varying the hyperparameters for each algorithm (such as number of trees and tree depth) and assessing the performance of the model using k-fold cross-validation \citep{1968Mosteller}. The scikit-learn implementation of this method is called \texttt{GridsearchCV} \footnote{\url{http://scikit-learn.org/stable/modules/generated/sklearn.model_selection.GridSearchCV.html}}. 

\begin{table}
\begin{center}
Hyperparameter Grid\\
\begin{tabular}{ | c | c |} 
\hline
\bf{n\_estimators} & 64, 128, 256, 512  \\ 
  \bf{max\_features} &1, 3, None  \\ 
  \bf{min\_samples\_leaf} & 1, 3, 10  \\ 
  \bf{criterion} & gini, entropy  \\
  \bf{min\_samples\_split} & 2, 3, 10 \\
  \bf{max\_depth} & 3, 6, 9, None \\
  \bf{learning\_rate} & 0.001, 0.01, 0.1, 0.5, 1.0 \\
 \hline
\end{tabular}
\caption{Hyperparameters for each machine learning algorithm (where applicable) which we explored during the gridsearch cross-validation.}
\label{tab:grids}
\end{center}
\end{table}

\begin{table}
\begin{center}
Hyperparameter Optimization Results (using \texttt{frames} features)\\
\begin{tabular}{ | c | c | c | c | c |} 
\hline
 & RF & ADA & EXT & GBT \\
\hline
\bf{n\_estimators} & 64 & 512 & 64 & 64  \\ 
  \bf{max\_features} & 3 & 1 & 1 & 1  \\ 
  \bf{min\_samples\_leaf} & 3 & 1 & 1 & 3  \\ 
  \bf{criterion} & gini & entropy & entropy & - \\
  \bf{min\_samples\_split} & 3 & 2 & 2 & 3 \\
  \bf{max\_depth} & None & 6 & None & 9 \\
  \bf{learning\_rate} & - & 1.0 & - & 0.1 \\
  \hline
  \bf{Mean Validation Score} & 0.974 & 0.975 & 0.974 & 0.974 \\
  \bf{Standard Deviation} & 0.004 & 0.003 & 0.004 & 0.002 \\
\end{tabular}
\caption{The most efficient variables for each machine learning method when only using the \texttt{frames} set of features.
\textbf{n\_estimators} is the number of trees, \textbf{max\_features} is the number of features to consider when looking for the best split within a tree, \textbf{min\_samples\_leaf} is the minimum number of objects required to be at a leaf node, \textbf{criterion} is the function that measures the quality of the split, \textbf{min\_samples\_split} is the minimum number of samples required to make a split, \textbf{max\_depth} limits the maximum depth of the trees, and \textbf{learning\_rate} (used only in the boosted model building methods of ADA and GBT) shrinks the contribution of each classifier by the set value. The \textbf{Mean Validation Score} is the accuracy which the best parameters achieved.} 
\label{tab:opt_res}
\end{center}
\end{table}

\begin{table}
\begin{center}
Hyperparameter Optimization Results (using 5 MINT features)\\
\begin{tabular}{ | c | c | c | c | c |}
\hline
 & RF & ADA & EXT & GBT \\
\hline
\bf{n\_estimators} & 64 & 512 & 64 & 256  \\ 
  \bf{max\_features} & 1 & 1 & 3 & 1  \\ 
  \bf{min\_samples\_leaf} & 1 & 3 & 3 & 10  \\ 
  \bf{criterion} & entropy & entropy & gini & - \\
  \bf{min\_samples\_split} & 10 & 3 & 3 & 10 \\
  \bf{max\_depth} & 3 & 4 & None & 9 \\
  \bf{learning\_rate} & - & 0.01 & - & 0.01 \\
  \hline
  \bf{Mean Validation Score} & 0.974 & 0.974 & 0.973 & 0.974 \\
  \bf{Standard Deviation} & 0.006 & 0.006 & 0.005 & 0.006 \\
\end{tabular}
\caption{The most efficient variables for each machine learning method when using 5 MINT selected features. Rows are as in Table \ref{tab:opt_res}.}
\label{tab:opt_res_MINT}
\end{center}
\end{table}

\begin{table}
\begin{center}
Hyperparameter Optimization Results (using all features)\\
\begin{tabular}{ | c | c | c | c | c |}
\hline
 & RF & ADA & EXT & GBT \\
\hline
\bf{n\_estimators} & 256 & 512 & 512 & 512  \\ 
  \bf{max\_features} & None & None & None & None  \\ 
  \bf{min\_samples\_leaf} & 1 & 1 & 1 & 10  \\ 
  \bf{criterion} & entropy & entropy & entropy & - \\
  \bf{min\_samples\_split} & 2 & 10 & 3 & 2 \\
  \bf{max\_depth} & None & 3 & None & 6 \\
  \bf{learning\_rate} & - & 0.1 & - & 0.1 \\
  \hline
  \bf{Mean Validation Score} & 0.979 & 0.980 & 0.981 & 0.981 \\
  \bf{Standard Deviation} & 0.003 & 0.004 & 0.004 & 0.004 \\
\end{tabular}
\caption{The most efficient variables for each machine learning method when using all available features in the sample. Rows are as in Table \ref{tab:opt_res}.}
\label{tab:opt_res_ALL}
\end{center}
\end{table}

The most efficient hyperparameters are listed in Tables \ref{tab:opt_res}, \ref{tab:opt_res_MINT}, and \ref{tab:opt_res_ALL} for the \texttt{frames} features test, the MINT selected features test, and the all features test respectively. The full grids can be seen in Table \ref{tab:grids}.

In most cases, 64 trees is an adequate number of estimators for all of the tested machine learning algorithms. However, it can be seen that the preferred trees are shallower when using five MINT selected features, yet the mean validation scores match or exceed that of the tests when using the \texttt{frames} set of features. This shows that MINT selected features do not degrade the predictive power, while reducing the number of computations.

\subsection{Using Random Forests as a motivation for improving \texttt{frames}}
\label{framesimprmeth}
Machine learning algorithms can also be used to optimise or check pre-existing decision boundaries such as the ones provided by the \texttt{frames} method in Equation \ref{eq:frames}. Perhaps a straight line very similar to the black dashed line in Figure \ref{fig:frames_cut} would be more accurate in classifying these objects.
To check if this is the case, we generated a Random Forest model on the training set, using only \texttt{PSFMAG\_I} and \texttt{CMODELMAG\_I} as input features (the same features as in the \texttt{frames} method for I-band). After performing a hyperparameter search (excluding \texttt{max\_features} as we only have 2 features), we then generated a fine grid of $x$ and $y$ coordinates spanning our training set magnitude limits and used the model to classify each of those points, which then outputs the decision boundary. We fit a straight line to the main trend of the decision boundary, and use this line instead of the one provided by the \texttt{frames} method of classification to classify objects, and determine if the Random Forest model can improve on it. We present the results of this test in Section \ref{sec:results_extratest}.

\section{Results}
\label{sec:results}
Presented in this section are the results from the tests described in previous sections. In particular we show results for the investigation into whether the clean flag generates artificial bias in the sample and model (Section \ref{sec:datprep}). We then compare the \texttt{frames} classification method with machine learning methods as introduced in Section \ref{sec:methods_ml}. We examine the use of Random Forests to improve the \texttt{frames} classification as discussed in Section \ref{framesimprmeth}, and finally present an example of multiclass classification where we classify objects as galaxies, stars, or QSOs. 

\subsection{Clean flag test} \label{sec:results_clean}
As described in Section \ref{sec:datprep}, we perform a Random Forest test without the pre-selection of objects labeled as clean in the CasJobs database, to assess how this affects accuracy.
Using the \texttt{frames} features defined in Section \ref{sec:data_fr}, with optimised Random Forest settings (after performing a new hyperparameter search because applying this flag changes the objects in the sample), the results from this test reach a total accuracy of 97.2\%. This is 0.2\% below the achievable rate when applying the \texttt{clean} flag.

As this work is essentially a proof of concept and not a comparison of machine learning models, we have chosen to utilise the \texttt{clean} flag in our tests to ensure the machine learning algorithm can build a model from reliable objects. This reduces noise in the model that could have influence on the placement of decision boundaries, which would cloud interpretability.

\subsection{Comparing \texttt{frames} and Machine Learning Methods} \label{sec:results_ml}
\begin{table*}
\begin{center}
\texttt{frames} method results (objc\_type vs template type using 1.5 million objects from the test sample.)\\
\begin{tabular}{ | c | c | c | c | c | c | c | c |} 
\hline
&  \multicolumn{2}{|c|}{Completeness} & \multicolumn{2}{|c|}{Purity} & \multicolumn{2}{|c|}{F1 Score} & Accuracy \\
 & Galaxies & Point Sources & Galaxies & Point Sources & Galaxies  & Point Sources \\
 \hline
 \bf{u} & 0.814 & 0.773 & 0.854 & 0.719 & 0.834 & 0.745 & 0.799\\
 \bf{g} & 0.957 & 0.937 &  0.961 &  0.930 &  0.959 & 0.933 & 0.949 \\
 \bf{r} & 0.990 & 0.932 &  0.959  & 0.983 & 0.974 & 0.957 & 0.968\\
 \bf{i} & 0.991 & 0.911 & 0.948  & 0.985 &  0.969 & 0.947 & 0.961 \\
 \bf{z} & 0.985 & 0.813 &  0.896  & 0.971 & 0.938 & 0.885 & 0.920 \\
 \hline
 \bf{ALL} & 0.986 & 0.943 &0.966 & 0.980 &0.977 & 0.961 & 0.971 \\
 \hline 
\end{tabular}
\caption{Results of classification for both galaxies and point sources using the \texttt{frames} method (Equation \ref{eq:frames}) in separate photometric filters, and using all filters. F1 score is the harmonic mean of the purity and completeness, and accuracy is the fraction of objects predicted correctly when comparing with the classification from fitted spectra. It is seen here that the \textbf{r} band filter gives the highest accuracy of classification, but when using a summation of  the fluxes from all of the photometric bands available, accuracy is increased.}
\label{tab:frames_res}
\end{center}
\end{table*}
In this section we make our main comparison between object classification using the SDSS \texttt{frames} criteria and the machine learning methods described in Section \ref{sec:methods_ml}.

We first assess object classification using the \texttt{frames} criteria (Equation \ref{eq:frames}). Table \ref{tab:frames_res} shows the results from the \texttt{frames} method of object classification in all filters separately, as well as combined. It is seen here by using all filters in combination that 97.1\% of object classifications match the classification given by spectroscopy. This result shows that the \texttt{frames} method performs object classification remarkably well while remaining simple and monotonic. The next tests will use machine learning methods to attempt to improve on this. 

\begin{table*}
\begin{center}
Machine Learning algorithm results (\texttt{frames} features)\\
\begin{tabular}{ | c | c | c | c | c | c | c | c |} 
\hline
&  \multicolumn{2}{|c|}{Completeness} & \multicolumn{2}{|c|}{Purity} & \multicolumn{2}{|c|}{F1 Score} & Accuracy \\
 & Galaxies & Point Sources & Galaxies & Point Sources & Galaxies  & Point Sources \\
 \hline
 \bf{Random Forest} & 0.986 & 0.955 & 0.973 & 0.976 &  0.979 & 0.966 & 0.974 \\
 \bf{Adaboost} & 0.985 & 0.954 &  0.972 & 0.975 & 0.979 & 0.965 & 0.973 \\
 \bf{ExtraTrees} & 0.986 & 0.953 & 0.972 & 0.977 & 0.979 & 0.965 & 0.974\\
 \bf{Gradient Boosted Trees} & 0.985 & 0.955 & 0.973 & 0.976 & 0.979 &  0.965 & 0.974\\
 \hline
\end{tabular}
\caption{Results of classification with four machine learning methods using the same features as in the \texttt{frames} method. Columns are as in Table \ref{tab:frames_res}.}
\label{tab:ML_res}
\end{center}
\end{table*}

Table \ref{tab:ML_res} shows the results of the different machine learning algorithms using the same set of features as the \texttt{frames} classification method. In all cases, the accuracy is slightly higher than that achieved by the \texttt{frames} method, with the average accuracy increase being 0.3\%, and the highest accuracy being 97.4\%.

\begin{table*}
\begin{center}
Machine Learning algorithm results (5 MINT selected features)\\
\begin{tabular}{ | c | c | c | c | c | c | c | c |} 
\hline
&  \multicolumn{2}{|c|}{Completeness} & \multicolumn{2}{|c|}{Purity} & \multicolumn{2}{|c|}{F1 Score} & Accuracy \\
 & Galaxies & Point Sources & Galaxies & Point Sources & Galaxies  & Point Sources \\
 \hline
 \bf{Random Forest} & 0.986 & 0.954 & 0.972 & 0.977 & 0.979 & 0.965 & 0.974\\
 \bf{Adaboost} & 0.986 & 0.953 & 0.971 &  0.977 & 0.979 & 0.965 & 0.974 \\
 \bf{ExtraTrees} &  0.986 & 0.956 & 0.973 & 0.976 & 0.979 & 0.966 & 0.974\\
 \bf{Gradient Boosted Trees} &  0.986 & 0.954 & 0.972 & 0.977 & 0.979 & 0.965 & 0.974\\
 \hline
\end{tabular}
\caption{Results of classification with four machine learning methods using 5 MINT selected features listed in Table \ref{tab:nMINT}. Columns are as in Table \ref{tab:frames_res}.}
\label{tab:ML_res_MINT5}
\end{center}
\end{table*}

Table \ref{tab:ML_res_MINT5} shows the results from the machine learning runs with 5 MINT selected features (see Table \ref{tab:nMINT} and Section \ref{subs:MINT}). The highest accuracy seen in this set of runs is also 97.4\%, showing that the MINT selected features are only as useful for classification accuracy as those selected for \texttt{frames} (except in the case of the Adaboost algorithm which shows a slight improvement of 0.1\%). It is of interest that there is only one feature in common between \texttt{frames} and MINT, and yet they succeed equally well under machine learning.

\begin{table*}
\begin{center}
Machine Learning algorithm results (all features)\\
\begin{tabular}{ | c | c | c | c | c | c | c | c |} 
\hline
&  \multicolumn{2}{|c|}{Completeness} & \multicolumn{2}{|c|}{Purity} & \multicolumn{2}{|c|}{F1 Score} & Accuracy \\
 & Galaxies & Point Sources & Galaxies & Point Sources & Galaxies  & Point Sources \\
 \hline
 \bf{Random Forest} & 0.990 & 0.964 & 0.978 & 0.983 & 0.984 & 0.973 & 0.980\\
 \bf{Adaboost} & 0.989 & 0.964 & 0.978 & 0.982 & 0.984 & 0.973 & 0.980  \\
 \bf{ExtraTrees} & 0.990 & 0.966 & 0.979 & 0.983 & 0.985 & 0.975 & 0.981\\
 \bf{Gradient Boosted Trees} &  0.989 & 0.968 & 0.980 & 0.982 &  0.985 & 0.975 & 0.981\\
 \hline
\end{tabular}
\caption{Results of classification for four machine learning methods using all available features in the catalogue. Columns are as in Table \ref{tab:frames_res}.}
\label{tab:ML_res_ALL}
\end{center}
\end{table*}

While using a low number of features (specially selected or not) in combination with machine learning methods yields good results, accuracy can be further improved by using as much data as possible. Table \ref{tab:ML_res_ALL} shows the results when using all available features in our catalogue, for each machine learning algorithm. It is seen here that the ExtraTrees and Gradient Boosted Trees method achieves the highest accuracies, correctly classifying 98.1\% of the objects in the test sample. 
This improves on the \texttt{frames} object classification accuracy by 1.0\%, which is $\approx33$\% improvement in the rate of misclassification. 

\subsection{Using Random Forests as a motivation for improving \texttt{frames}} \label{sec:results_extratest}

\begin{figure}
\includegraphics[width=\columnwidth]{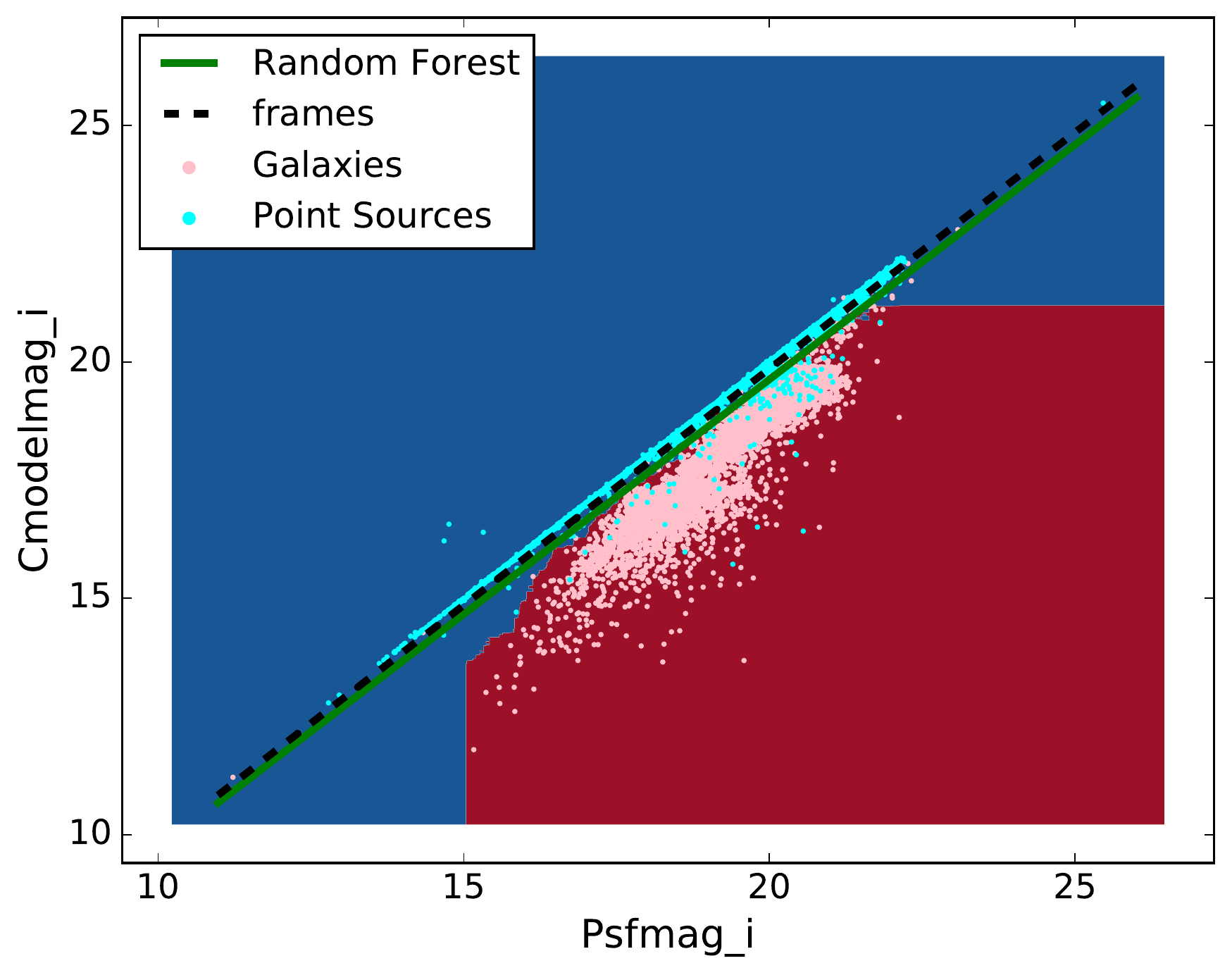}
\caption{The decision boundaries generated by a Random Forest run using \texttt{PSFMAG\_I} and \texttt{CMODELMAG\_I} as features. The training data has been overplotted with colour denoting spectroscopic classification (colours as in Figure \ref{fig:db_cut}). The original frames method of classification is shown by the black dashed line, and the Random Forest motivated method of classification is shown by the green line.}
\label{fig:framesimpr}
\end{figure}

In Section \ref{framesimprmeth}, we discussed how Random Forests could be used to check or optimise a method like \texttt{frames}.
Figure \ref{fig:framesimpr} shows that by fitting a line to the main trend of the decision boundary used by the Random Forest model, we obtain a slightly shallower line than the one given by the \texttt{frames} method, with the equation being $y = 0.993x + -0.218$.
Using this new line to classify the test data, we improve the accuracy of object classification in the I band by $\approx$0.8\%, and discover that objects are more likely to be point sources when \texttt{CMODELMAG\_I} is lower than \texttt{PSFMAG\_I} at fainter magnitudes (though this effect decreases as brightness of the object increases). 

\subsection{Multiclass Classification}

The SDSS pipeline outputs both a classification type and subtype from the template fitting of spectra (e.g. type = point source, sub type = star or QSO). Therefore, it is possible to test machine learning algorithms with the more complex task of deciding between more classifications than just galaxy or point source.

\begin{figure}
\includegraphics[width=\columnwidth]{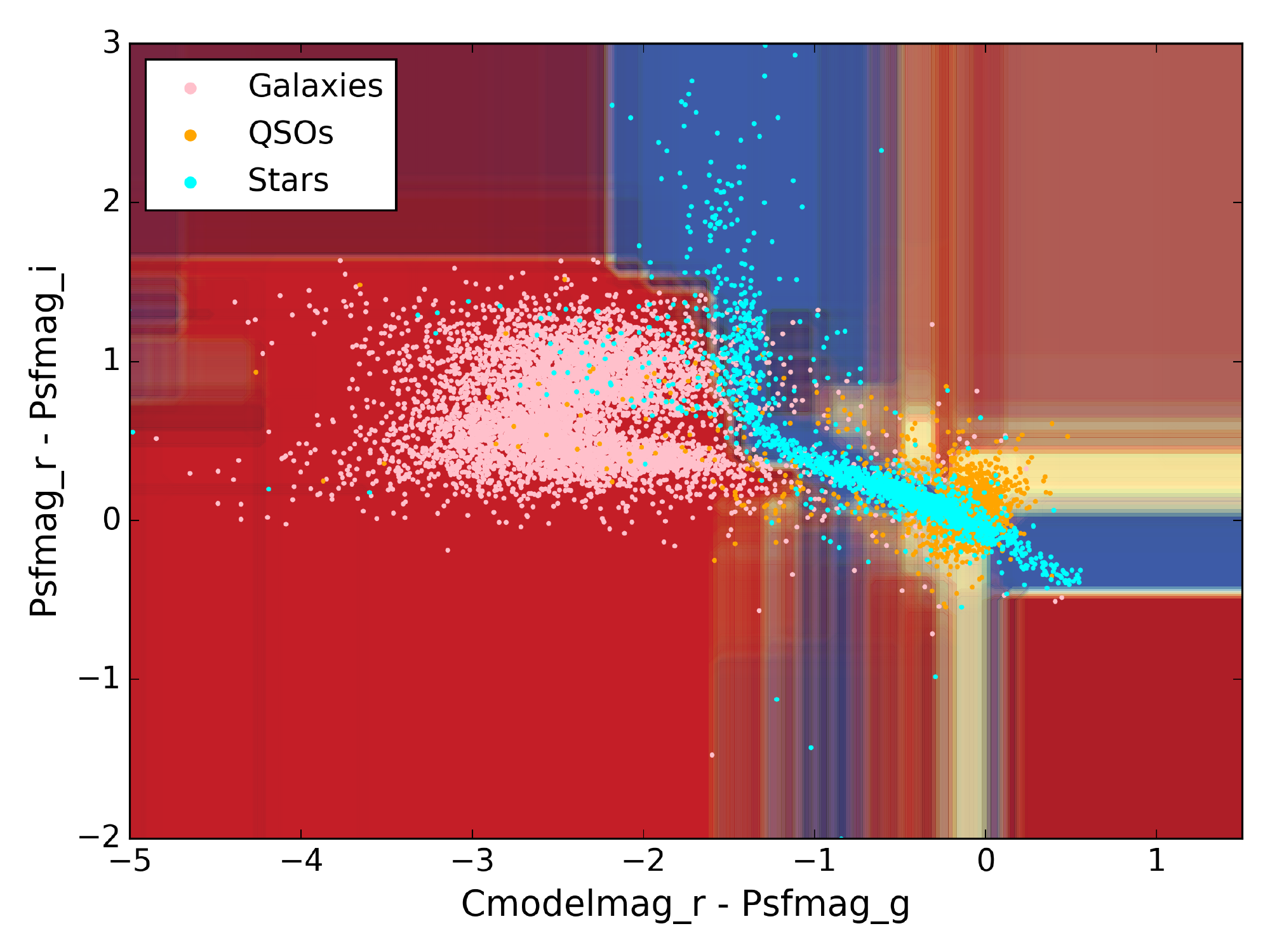}
\caption{The decision boundaries  generated by an example Random Forest run on a multiclass problem using two photometric colours as features. The training data has been overplotted with colour denoting spectroscopic classification. The colours are as in Figure \ref{fig:db_cut}, with point sources being split into stars (cyan points, blue decision boundaries) and QSOs (orange points, yellow decision boundaries).}
\label{fig:multiclass}
\end{figure}

Figure \ref{fig:multiclass} shows the decision boundaries from a Random Forest run using two photometric colours where the algorithm was asked to decide if an object was a star, galaxy, or QSO - a multiclass problem. The two colours were chosen as features for this example because they better disperse the data than using two \texttt{frames} features. The training process is the same as for a binary classification problem except that here the decision trees in the forest will have a fraction of leaves which identify QSOs. After a fresh hyperparameter search, we find the Random Forest achieves an object classification accuracy of 89.6\%. This accuracy is lower than in previous tests due to the model's inability to accurately distinguish between stars and QSOs; this may be due to their inherent similarities as point sources. Nevertheless, this example points towards the potential of this paper's ML methods for more extensive multiclassification problems. 

\section{Conclusion}
\label{sec:discussion}

Research into the area of machine learning has become prevalent in recent years, and it is important that research fields such as astronomy rapidly benefit from new modelling methods. 

This paper has showcased novel machine learning methods by revisiting the long standing object classification method used in the SDSS pipeline, \texttt{frames}, with the aim of increasing object classification accuracy using photometric data. We have developed a pipeline that offers in-depth analysis of machine learning models using \textit{treeinterpreter}, which has the ability to select the most important and relevant features specific to the input data using MINT. In practice, the pipeline improves on the \texttt{frames} object classification accuracy by 1.0\%, which is $\approx33$\% improvement in the rate of misclassification. 

It can be seen from the results that while the \texttt{frames} method of classification performs very well, machine learning methods (especially feature driven and tuned models) can outperform them. Indeed, there are several reasons for considering methods such as those outlined in this paper. 

Firstly, it has been shown that tree based methods offer at least some level of interpretability. Machine learning models and feature selection methods such as MINT may choose to use features that do not seem to be obvious, so figuring out how and why the model is working has been difficult. With new codes such as \textit{treeinterpreter}, we have shown that the models can be analyzed in such a way as to provide insight into which features are important to the problem and why. Using such methods, it is possible for the machine to pick out relations/correlations that have been previously missed. 

Secondly, a higher degree of classification accuracy can be achieved - one closer to that obtained by fitting spectra. The machine learning algorithms also output probabilities for each classification, allowing users to single out objects which are a problem for the machine learning model.

Thirdly, the machine learning method of classification is computationally almost as quick as the \texttt{frames} method. For future surveys, speed of data processing will become a very important problem. Our method could be included in the pipeline of a new survey, where a standard training set is created and given to the pipeline (from science verification data for example), and the model could be continuously improved as new data is observed. 

This work is an example of how new methods like \textit{treeinterpreter} and MINT are useful in understanding the relationship between data and the performance of machine learning models. This analysis would have to repeated for new datasets from different astronomical surveys because the results presented here are not trivially transferable. In the future, as well as being incorporated into survey data processing pipelines, these methods could be applied to other problems in astronomy such as predicting redshifts or the physical properties of galaxies, and offer new insights into how and why machine learning algorithms make their decisions.

\section*{Acknowledgements}
XMA would like to thank the Institute of Cosmology and Gravitation (ICG), the University of Portsmouth, and SEPNet for funding.
XMA would like to thank Claire Le Cras for her helpful discussion, patience, and support.
XMA would also like to thank Gary Burton for his help when using the Sciama HPC cluster.

DB is supported by STFC consolidated grant ST/N000668/1.

Numerical computations were done on the Sciama High Performance Compute (HPC) cluster which is supported by 
the ICG, SEPNet and the University of Portsmouth.

Funding for SDSS-III has been provided by the Alfred P. Sloan Foundation, the Participating Institutions, the National Science Foundation, and the U.S. Department of Energy Office of Science. The SDSS-III web site is http://www.sdss3.org/.

SDSS-III is managed by the Astrophysical Research Consortium for the Participating Institutions of the SDSS-III Collaboration including the University of Arizona, the Brazilian Participation Group, Brookhaven National Laboratory, Carnegie Mellon University, University of Florida, the French Participation Group, the German Participation Group, Harvard University, the Instituto de Astrofisica de Canarias, the Michigan State/Notre Dame/JINA Participation Group, Johns Hopkins University, Lawrence Berkeley National Laboratory, Max Planck Institute for Astrophysics, Max Planck Institute for Extraterrestrial Physics, New Mexico State University, New York University, Ohio State University, Pennsylvania State University, University of Portsmouth, Princeton University, the Spanish Participation Group, University of Tokyo, University of Utah, Vanderbilt University, University of Virginia, University of Washington, and Yale University.

%%%%%%%%%%%%%%%%%%%%%%%%%%%%%%%%%%%%%%%%%%%%%%%%%%

%%%%%%%%%%%%%%%%%%%% REFERENCES %%%%%%%%%%%%%%%%%%

\bibliographystyle{mnras}
\bibliography{references}

%%%%%%%%%%%%%%%%%%%%%%%%%%%%%%%%%%%%%%%%%%%%%%%%%%

%%%%%%%%%%%%%%%%% APPENDICES %%%%%%%%%%%%%%%%%%%%%

\appendix
\section{Casjobs SQL Query}
\label{app:sql}
This is the SQL Query submitted to Casjobs to obtain all the values required to calculate the whole sample used in this work.\\
\texttt{select s.specObjID, s.class  as spec\_class, q.objid,\\
q.dered\_u,q.dered\_g,q.dered\_r,q.dered\_i,
q.dered\_z,\\
q.modelMagErr\_u,q.modelMagErr\_g, q.modelMagErr\_r,
q.modelMagErr\_i,q.modelMagErr\_z,
q.extinction\_u,q.extinction\_g,q.extinction\_r,
q.extinction\_i,q.extinction\_z,
q.cModelMag\_u,q.cModelMagErr\_u,
q.cModelMag\_g,q.cModelMagErr\_g,
q.cModelMag\_r,q.cModelMagErr\_r,
q.cModelMag\_i,q.cModelMagErr\_i,
q.cModelMag\_z,q.cModelMagErr\_z,
q.psfMag\_u,q.psfMagErr\_u, 
q.psfMag\_g,q.psfMagErr\_g, 
q.psfMag\_r,q.psfMagErr\_r, 
q.psfMag\_i,q.psfMagErr\_i, 
q.psfMag\_z,q.psfMagErr\_z,\\
q.fiberMag\_u,q.fiberMagErr\_u,
q.fiberMag\_g,q.fiberMagErr\_g,
q.fiberMag\_r,q.fiberMagErr\_r,
q.fiberMag\_i,q.fiberMagErr\_i,
q.fiberMag\_z,q.fiberMagErr\_z,
q.expRad\_u,
q.expRad\_g,
q.expRad\_r,
q.expRad\_i,
q.expRad\_z,
q.clean,
s.zWarning\\
\\
into mydb.specPhotoDR12 from SpecObjAll as s 
join photoObjAll as q on s.bestobjid=q.objid
left outer join Photoz as p on s.bestobjid=p.objid}

%%%%%%%%%%%%%%%%%%%%%%%%%%%%%%%%%%%%%%%%%%%%%%%%%%

% Don't change these lines
\bsp	% typesetting comment
\label{lastpage}
\end{document}